# $^{39}$Ar Detection at the $10^{-16}$ Isotopic Abundance Level with Atom Trap Trace Analysis

W. Jiang,[1] W. D. Williams,[1] K. Bailey,[1] A.M. Davis,[2,3] S.-M. Hu,[4] Z.-T. Lu,[1,2,5] T.P. O'Connor,[1] R. Purtschert,[6] N.C. Sturchio,[7] Y.R. Sun,[4] and P. Mueller[1]

[1]*Physics Division, Argonne National Laboratory, Argonne, Illinois 60439, USA*

[2]*Enrico Fermi Institute, University of Chicago, Chicago, Illinois 60637, USA*

[3]*Department of Geophysical Sciences, University of Chicago, Chicago, Illinois 60637, USA*

[4]*Hefei National Laboratory for Physical Sciences at the Microscale, University of Science and Technology of China, Hefei, Anhui 230026, China*

[5]*Department of Physics, University of Chicago, Chicago, Illinois 60637, USA*

[6]*Climate and Environmental Physics, University of Bern, CH-3012 Bern, Switzerland*

[7]*Department of Earth and Environmental Sciences, University of Illinois at Chicago, Chicago, Illinois 60607, USA*

Atom Trap Trace Analysis (ATTA), a laser-based atom counting method, has been applied to analyze atmospheric $^{39}$Ar (half-life = 269 yr), a cosmogenic isotope with an isotopic abundance of $8\times10^{-16}$. In addition to the superior selectivity demonstrated in this work, counting rate and efficiency of ATTA have been improved by two orders of magnitude over prior results. Significant applications of this new analytical capability lie in radioisotope dating of ice and water samples and in the development of dark matter detectors.

Ultrasensitive trace analysis of radioactive isotopes has enabled a wide range of applications in both fundamental and applied sciences. The three long-lived noble-gas isotopes, $^{85}$Kr (half-live = 10.8 yr), $^{39}$Ar and $^{81}$Kr (230,000 yr), are particularly significant for applications in the earth sciences [1]. Being immune to chemical reactions, these three isotopes are predominantly stored in the atmosphere, they follow relatively simple mixing and transport processes in the environment, and they can be easily extracted from a large quantity of water or ice samples. Indeed they possess ideal geophysical and geochemical properties for radioisotope dating.

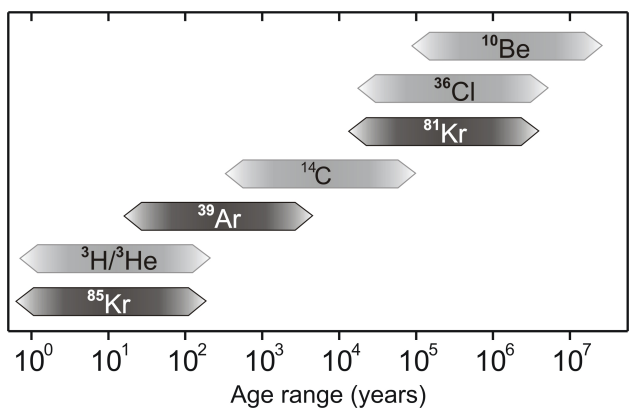

**FIG. 1.** Dating ranges of $^{85}$Kr, $^{39}$Ar, $^{81}$Kr and other established radioisotope tracers. At an age much shorter than the half-life, the variation of the isotopic abundance becomes too small to measure accurately. On the other hand, when the age is much longer than the half-life, the abundance itself becomes too small to measure accurately. The range shown here is approximately 0.1 – 10 times the half-life.

Dating ranges of radioisotope tracers follow closely their radioactive half-lives (see Fig. 1). The half-lives of the three noble gas isotopes have different orders of magnitude, allowing them to cover a wide range of ages. In particular, $^{39}$Ar conveniently fills an apparent gap between $^{85}$Kr, $^3$H/$^3$He on the shorter side and $^{14}$C on the longer side. This makes $^{39}$Ar a much desired isotope for dating environmental samples on the time scale of a few hundred years [2]. For example, $^{39}$Ar dating in combination with $^{14}$C dating can be used to study mixing processes in ocean water or groundwater, with implications for modeling global and regional climate changes [3][4].

On a different front, ultrasensitive detection of argon and krypton isotopes is critical for developing the next-generation dark matter detectors based on liquid noble gases [5][6]. Here, $^{39}$Ar and $^{85}$Kr are major sources of unwanted radioactive background. In particular, finding large underground sources of argon gas depleted in $^{39}$Ar would be key in realizing the full potential of liquid argon based detectors [7].

$^{39}$Ar is produced in the atmosphere through cosmic-ray induced nuclear reactions and equilibrates at an isotopic abundance of $8\times10^{-16}$ [8], a level so low that it poses an extreme challenge to analytical methods. Only two previous methods, Low-Level Decay Counting (LLC) and Accelerator Mass Spectrometry (AMS), have succeeded in demonstrating $^{39}$Ar analysis of environmental samples. In LLC, proportional gas counters located in an underground laboratory at the University of Bern, Switzerland, are employed to detect the β-decay of $^{39}$Ar [8]. A typical measurement requires 0.3-1 liter of argon gas at standard temperature and pressure (STP) extracted from 1-3 tons of water, and takes 8-60 days [9]. The detection efficiency of $^{39}$Ar LLC lies in the range of $10^{-4}$, while its detection limit of $^{39}$Ar/Ar ~ $4\times10^{-17}$ is governed by variations of environmental radioactivity background. For AMS measurements, the Argonne ATLAS heavy ion accelerator was used in combination with a gas-filled magnetic spectrograph to selectively detect the high energy ions [10]. By counting atoms instead of decays, AMS is much faster and more efficient than LLC: an AMS analysis of $^{39}$Ar takes approximately 8 hours, requires 2 mL STP of argon gas, and reaches an efficiency of ~$2.8\times10^{-3}$ and a $^{39}$Ar/Ar detection limit of $4.3\times10^{-17}$. However, further application of

this technique has been difficult due to its dependence on the use of a large accelerator, of which access (beam-time) is limited.

Atom Trap Trace Analysis (ATTA) is a laser-based atom counting method [11]. Its apparatus consists of lasers and vacuum systems of table-top size. At its center is a magneto-optical trap (MOT) to capture atoms of the desired isotope using laser beams. A sensitive CCD camera detects the laser induced fluorescence emitted by the atoms held in vacuum. Trapping force and fluorescence detection require the atom to repeatedly scatter photons at a high rate ($\sim 10^7$ s$^{-1}$). This is the key to the superior selectivity of ATTA because it only occurs when the laser frequency precisely matches the resonance frequency of a particular atomic transition. Even the small changes in the atomic transition frequency between isotopes of the same element – the so called isotope shifts caused by changes in nuclear mass and moments – are sufficient to perfectly distinguish between the isotopes. ATTA is unique among trace analysis techniques as it is virtually free of interferences from other isotopes, isobars, or molecular species. Previously, ATTA has been developed to analyze $^{85}$Kr and $^{81}$Kr at the $10^{-12}$ isotopic abundance level [11][12][13], and has been used to perform $^{81}$Kr dating of old groundwater samples [14].

In this work, by detecting $^{39}$Ar in atmospheric samples we demonstrate that ATTA can perform rare isotope analysis at and below the $10^{-15}$ abundance level. Laser cooling and trapping of stable argon isotopes based on the 4s(3/2)$_2$ - 4p(5/2)$_3$ cycling transition has previously been achieved for various atomic physics investigations [15][16]. Development towards an atom-trap based instrument for $^{39}$Ar analysis is also pursued by Welte *et al.* [17]. Isotope shift and hyperfine structure of $^{39}$Ar for this transition have been measured recently [18][19].

The basic layout of the atomic beam apparatus used for this work is described in detail in [13]. In short, a radio-frequency (RF) driven, inductively-coupled gas discharge source produces a beam of metastable argon atoms in the 4s(3/2)$_2$ level. Subsequently, a transverse laser cooling and focusing stage and a Zeeman slower collimate and cool the atomic beam for efficient transfer into the MOT. A mechanical beam chopper periodically blocks the atomic beam during the single-atom detection period to reduce photon background caused by laser induced fluorescence from the atomic beam. Fluorescence light emitted by the trapped atoms is imaged onto a sensitive CCD camera for atom detection.

A laser system based on an extended cavity diode lasers and semiconductor power amplifiers supplies the required narrowband laser light (~300 kHz linewidth) at 812 nm for all laser cooling stages. The frequency of the laser light is referenced to $^{40}$Ar via saturated absorption spectroscopy in a discharge gas cell. Acousto-optical modulators provide the relative frequency shifts required to trap $^{38}$Ar and $^{39}$Ar. Additional sidebands for hyperfine repumping are required for trapping of $^{39}$Ar (nuclear spin = 7/2), because the trapping light on resonance with the F = 11/2 - 13/2 hyperfine transition has a finite probability to also excite to the F = 9/2 and 11/2 states in the upper level. Subsequent decay to the F = 7/2 or 9/2 states in the lower level would take the atom out of the cooling cycle and lead to atom loss. To counteract this depopulation, we added laser light resonant with all four additional ΔF = +1 transitions to optically pump the atoms back to the F = 11/2 state in the lower level.

Compared with the previous ATTA instrument demonstrated for krypton isotope analysis [13], a factor of 200 increase in both capture efficiency and counting rate have been realized with the following upgrades: (a) the gas discharge source is cooled with liquid nitrogen to decrease the velocity of the atomic beam; (b) the length of the transverse cooling stage has been doubled to 20 cm to increase the transverse capture velocity; (c) the laser system has been upgraded with additional semiconductor tapered amplifiers to provide a total of 2 W of laser light at 812 nm; and (d) an atomic beam focusing stage has been added to further improve the solid angle overlap of the atomic beam with the MOT volume.

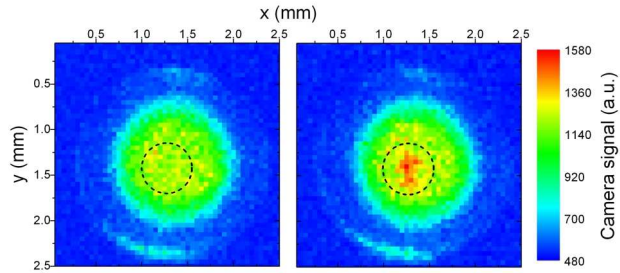

**Fig. 2.** False-color CCD images of the MOT at $^{39}$Ar settings. In (a) no atom is present; the background is dominated by scattered laser light imaged through a 1 mm diameter aperture and an additional ring-shaped diffraction pattern. In (b) the fluorescence emitted by a single $^{39}$Ar atom can be clearly distinguished above background. The dashed circle indicates a 0.5 mm diameter region of interest for averaging the CCD signal.

Prior to $^{39}$Ar trapping, we optimized the trap system with $^{38}$Ar (isotopic abundance = 0.063%) achieving capture rates of $1.3 \times 10^9$ atoms per second as measured from the fluorescence signal of the large $^{38}$Ar atom cloud. For $^{39}$Ar detection, the trap system alternated between a mode for optimum capture and a mode for single atom detection that minimizes background from scattered light, following a 350 ms *vs.* 100 ms cycle, respectively. During the detection mode a CCD camera recorded an image of the MOT center. Atom fluorescence signals were obtained from averaging the image brightness within a circular region-of-interest that was matched to the spatial distribution of the atom cloud. Sample CCD images with and without an $^{39}$Ar atom are shown in Fig. 2.

Optimization of the single atom signal-to-noise level was performed with $^{38}$Ar by reducing the loading rate to roughly one atom per second. Typical $^{38}$Ar single atom signals are shown in Fig. 3(a) and 3(b). They display discrete steps representing the fluorescence signal from individual $^{38}$Ar atoms. The single atom detection threshold with optimum signal-to-noise ratio was determined to be 5.5$\sigma$ above background based on an analysis of $^{38,39}$Ar single atom signal sizes and of the random detector noise under $^{39}$Ar detection conditions.

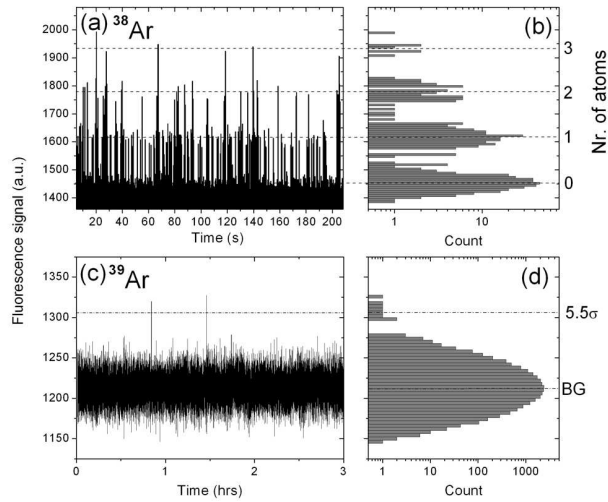

**Fig. 3.** Laser induced fluorescence signals from single $^{38}$Ar and $^{39}$Ar atoms. (a) $^{38}$Ar signals vs. time with the atom loading rate reduced to about one per second. (b) The histogram reveals discrete signal levels corresponding to the fluorescence emitted from individual $^{38}$Ar atoms. (c) $^{39}$Ar signals vs. time. Two $^{39}$Ar atoms were detected with >5.5$\sigma$ significance and remained in the trap for a total of eight detection cycles as seen in the respective histogram in (d).

In order to demonstrate $^{39}$Ar trapping from an atmospheric sample we alternated among three frequencies for the trapping laser and sidebands: -20 MHz, -6 MHz, and +10 MHz relative to the transitions in $^{39}$Ar. All other experimental parameters were held constant. Laser cooling and trapping of $^{39}$Ar should only occur at -6 MHz, while the other two frequencies serve as background measurements. We recorded fluorescence signals at each frequency for three hours. At the beginning and the end of each three-hour period, we checked the system by measuring $^{38}$Ar loading rate and single atom signal. The full cycle of measurements at all three frequencies was repeated a total of 19 times. For atmospheric samples, we used commercial argon gas, which had been extracted from air within the preceding year.

The data from a three-hour run at -6 MHz is illustrated in Fig. 3(c) and 3(d): two events stand out above background and are interpreted as fluorescence signals from individual trapped $^{39}$Ar atoms. On average, we obtained a signal-to-noise ratio of 7 for $^{39}$Ar single-atom signals in 100 ms integration time. The CCD camera image shown in Fig. 2(b) corresponds to data collected from the second event in Fig. 3(c). The spatial distribution of the atom fluorescence of $^{39}$Ar matches that of $^{38}$Ar. Zooming into the data in Fig. 3(c) also reveals that the fluorescence signals persisted over several loading-detection cycles, indicating that the $^{39}$Ar atoms stayed in the trap for more than 450 ms. Accordingly, consecutive detection cycles above threshold are counted only once. In the histogram in Fig. 3(d) a total of eight cycles originating from two atoms lie above the detection threshold.

In total, we registered twelve events above the 5.5σ detection threshold at -6 MHz, while no such events were observed at either -20 MHz or +10 MHz as shown in Fig. 4(a). Assuming a Gaussian distribution of the camera signal background, there is less than 1% probability for one out of the twelve events to be caused by random background. Based on the zero count at -20 MHz and +10 MHz we can exclude contaminations due to other atomic or molecular species above $1\times10^{-16}$ at the 90% confidence level. To illustrate the expected line shape we plotted the normalized atom count rate of $^{81}$Kr from an atmospheric sample ($^{81}$Kr/Kr = $5\times10^{-13}$) as a dashed line in Fig. 4(a) on the same frequency scale relative to the cycling transition. The $^{81}$Kr data were taken with the same apparatus at finer laser frequency intervals. Its line shape should mirror that of $^{39}$Ar due to the nearly identical line strength and linewidth of the corresponding transition in $^{81}$Kr. Because of its higher isotopic abundance, the maximum $^{81}$Kr count rate was 420 atoms per hour. In summary, the unambiguous identification of $^{39}$Ar atom counts is based on their distinct dependence on the laser frequency, the high signal-to-noise ratio and the specific spatial characteristic of the fluorescence signal.

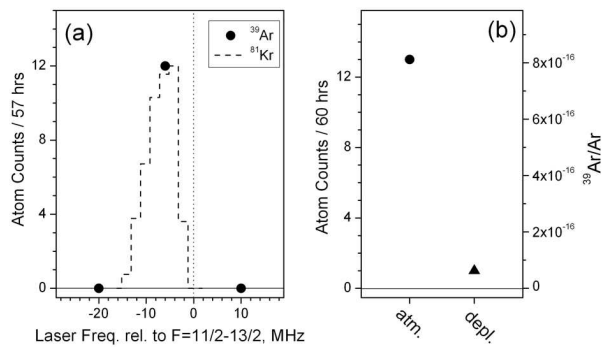

**Fig. 4.** (a) $^{39}$Ar atom counts as a function of laser frequency. Atoms were only detected at a laser frequency of -6 MHz relative to the cycling transition. The dashed line represents the normalized loading rate of $^{81}$Kr on the same relative frequency scale for comparison. (b) $^{39}$Ar counts per 60 hours at -6 MHz for the atmospheric and depleted sample.

In addition to the atmospheric sample, we analyzed a sample depleted in $^{39}$Ar. The sample was measured by LLC to contain less than 7% of the atmospheric $^{39}$Ar abundance at the 86% confidence level. In this measurement we applied a partial gas recirculation scheme [13], where a large fraction (~99%) of the sample entering the atomic beam apparatus was repeatedly pumped back to the source. This reduced the sample consumption rate to 0.45 mL STP per hour. We note that this rate can be further reduced, and the detection efficiency increased, by an order of magnitude with a full recirculation scheme [13].

In order to measure the $^{39}$Ar abundance ratio between the depleted and atmospheric sample, we recorded atom counts at -6 MHz in eight 15-hour segments, alternating between the two samples. The stability of the instrument was checked with $^{38}$Ar measurements every three hours. The outgassing rate of atmospheric argon into the vacuum system was determined to be less than 0.5% of the sample consumption rate, leading to a negligible cross sample contamination. We counted a total of 13 $^{39}$Ar atoms for the atmospheric sample, while only one $^{39}$Ar atom was recorded for the depleted sample (Fig. 4(b)). Thus the $^{39}$Ar abundance in the depleted sample was determined to be $8^{+18}_{-7}$% of the atmospheric value in good agreement with the LLC measurement. The $^{39}$Ar counting efficiency with partial gas recirculation was $2.2(6) \times 10^{-5}$.

Based on the average $^{38}$Ar loading rate of $1.3 \times 10^9$ atoms per second one might expect a $^{39}$Ar loading rate of up to 6 atoms per hour for the atmospheric sample, a factor of ~30 more than what we observed. While the causes of the reduced loading efficiency for $^{39}$Ar are not fully understood, we ascribe it primarily to two effects: (a) the switching between loading and detection mode for $^{39}$Ar causes a fraction of atoms to be lost from the trap before they are detected due to the finite trap lifetime; and (b) incomplete hyperfine pumping of $^{39}$Ar in any of the laser cooling steps results in leakage to hyperfine states not in resonance with the laser light. Therefore, we anticipate two immediate steps to improve the $^{39}$Ar loading rate: reduction of the scattered light level to enable continuous monitoring of the $^{39}$Ar loading, and further optimization of our hyperfine pumping scheme in terms of frequency, intensity, and alignment of the laser light. These improvements of the trapping system would raise both counting rate and counting efficiency. Being limited only by counting statistics rather than unspecific background events, this would enable us to lower the detection limit in the near future with the goal of reaching isotopic abundance sensitivities at the 10$^-$ [18] level. With this new analytical tool exciting applications for $^{39}$Ar trace analysis in earth sciences and fundamental physics will become possible.

We thank Y. Ding for his extensive contribution during the early stage of this project. We also thank C.-F. Cheng, J.P. Greene and R.J. Holt for their contributions to this work. This work is supported by the Department of Energy, Office of Nuclear Physics, under contract DEAC02-06CH11357; and by National Science Foundation, Division of Earth Sciences, under Award No. EAR-0651161. S.-M. Hu acknowledges support from National Natural Science Foundation of China (90921006).